\documentclass[12pt,preprint]{aastex}

\begin{document}
\title{Kinematic Density Waves in Accretion Disks}
\author{Svetlin V. Tassev}
\affil{Harvard-Smithsonian Center for Astrophysics, 60 Garden
St.,Cambridge, MA 02138} 
\email {stassev@cfa.harvard.edu} 
\and
\author{Edmund Bertschinger}
\affil{Department of Physics and Kavli Institute for Astrophysics
and Space Research, MIT Room 37-602A, 77 Massachusetts Ave.,
  Cambridge, MA 02139}
\email{edbert@mit.edu}

\begin{abstract}
When thin accretion disks around black holes are perturbed, the main
restoring force is gravity.  If gas pressure, magnetic stresses, and
radiation pressure are neglected, the disk remains thin as long as
orbits do not intersect.  Intersections would result in pressure
forces which limit the growth of perturbations.  We find that a
discrete set of perturbations is possible for which orbits remain
non-intersecting for arbitrarily long times.  These modes define a
discrete set of frequencies.  We classify all long-lived
perturbations for arbitrary potentials and show how their mode
frequencies are related to pattern speeds computed from the
azimuthal and epicyclic frequencies.  We show that modes are
concentrated near radii where the pattern speed has vanishing radial
derivative.

We explore these modes around Kerr black holes as a possible
explanation for the high-frequency quasi-periodic oscillations of
black hole binaries such as GRO J1655-40.  The long-lived modes are
shown to coincide with diskoseismic waves in the limit of small
sound speed.  While the waves have long lifetime, they have the
wrong frequencies to explain the pairs of high-frequency
quasi-periodic oscillations observed in black hole binaries.
\end{abstract}

\section{Introduction}

X-ray emission from accretion disks around compact objects often
shows high-frequency flickering in the form of quasi-periodic
oscillations (QPO).  Several galactic binaries containing black
holes show pairs of QPO of fixed, relatively high frequencies having
ratios very close to 3:2 \citep{Remillard2006}.  These include GRO
J1655-40 with 300 and 450 Hz QPO \citep{Strohmayer2001}; XTE
J1550-564 with 184 and 276 Hz \citep{Miller2001}; GRS 1915+105 with
113 and 168 Hz \citep{Remillard2004}; and H1743-322 with 165 and 240
Hz \citep{Homan2005}.

The origin of these quasi-periodic oscillations is still unclear.
One possibility is that orbiting blobs or hot spots follow
non-circular orbits whose epicyclic frequencies are related to the
observed frequencies (Stella, Vietri \& Morsink 1999).  This model
has been investigated in detail by \cite{Schnittman2004} and
\cite{Schnittman2005}, who use a ray-tracing code to model the power
spectrum of light curve variations.  The model has two serious
shortcomings.  First, special orbits must be chosen by hand (e.g.,
the radii of the nearly circular orbits) to produce the correct
frequencies, so this model lacks predictive power.  Moreover, the
blobs can persist only a few rotation periods before being sheared
apart by differential rotation.

Another model that can reproduce a 3:2 frequency ratio is the
relativistic torus model of \cite{Rezzolla2003}.  To produce the 3:2
ratio, the model requires a black hole spin close to maximal. The
actual value of frequency depends on the radial extent of the torus.
This is a free parameter of the model, which therefore also lacks
predictive power.

The fixed frequencies and 3:2 ratio of these QPO pairs suggests that
they may represent nonlinearly driven resonant frequencies in the
curved spacetime around the black hole (Abramowicz \& Klu{\'z}niak
2001; Rebusco 2004; Lee, Abramowicz \& Klu{\'z}niak 2004).  This
idea has been studied using a mathematical model for parametric
resonance but has not yet been shown to arise automatically from
nonlinear dynamics in general relativity.

Another possible explanation for the fixed frequencies of
high-frequency QPO is diskoseismic waves similar to helioseismic
$g$-modes \citep{Wagoner2001}.  This model has the advantage of
predicting a discrete set of frequencies.  However, a frequency
ratio of 3:2 only obtains for a few particular values of the black
hole spin that are unlikely to obtain for all observed systems.
Thus, similarly to the hot spot model, this model requires tuning of
parameters.

The models can be classified as to whether the oscillating
structures are particle-like (blobs, with or without resonant
interactions) or wave-like (oscillating tori, diskoseismic waves).
Blob models are in many respects the simplest, as the only important
physics is gravity.  It is natural to ask whether a similarly simple
description is possible for waves.

In this paper we show that waves in an accretion disk can also be
described using simple kinematics in curved spacetime. We
investigate nearly kinematic density waves, which are waves of
compression and rarefaction of particles freely orbiting in a disk.
These density waves are very similar to the density waves which
underlie the Lin-Shu hypothesis of galaxy spiral structure, apart
from several important differences. Our model neglects self-gravity,
pressure, magnetic stresses, and all forces except the gravity of
the central compact object. A kinematic model of perturbations in thin accretion disks around Kerr black holes was considered in \cite{Amin2006}. However, they allow for fluid streamlines to intersect forming shocks, as we will describe below.

Early models for galactic spiral structure (Lindblad 1957, Kalnajs
1973) failed to predict sufficiently long-lived patterns because
differential rotation relatively quickly winds up material arms;
moreover, these models neglected self-gravity which is invalid for
galactic disks (Toomre 1977; Binney \& Tremaine 1987).  However,
accretion disks around black holes contain relatively much less mass
than their central ``bulge'', so self-gravity is much less
important.  If other forces are unimportant --- as might be expected
for a thin disk --- then the remaining challenge is to prevent the
waves from becoming tightly wound in a short time.  This is exactly
the problem considered in the galactic context by
\cite{Kalnajs1973}.  Our model is a generalization of his kinematic
density waves to arbitrary potentials and to the strongly curved
spacetime around a Kerr black hole.

We choose to investigate a thin disk, since it is analytically
tractable. In this case pressure forces are not important, otherwise
the disk would not remain very thin.  For consistency, shocks must
not form in the gas.  Thus the perturbations must evolve so that
fluid streamlines avoid intersections.  The boundary between regions
where streamlines cross and where they do not is a caustic surface.
Our model attempts to find long-lived kinematic density waves free
from caustics.

\section{Disk Perturbations in Generic Potentials}

In order to study the dynamics of thin accretion disks, we will
perturb an initially stationary and axisymmetric thin disk and then
impose the condition that particle trajectories do not intersect.
The surface number density is $n_0(\bar r)$ where $\bar r$ is the
radius of an unperturbed orbit.  This assumes that the effective
potential is axisymmetric and each orbit has a conserved angular
momentum as well as energy.

Perturbations are added using the Lagrangian framework based on
individual particle trajectories.  Thus, at each instant of time,
there exists a nonlinear mapping that takes the coordinates of a
particle in the unperturbed disk to the coordinates of the same
particle in a perturbed disk.  We choose to map the Lagrangian
coordinates (which we choose to call $\bar r$ and $\bar\phi$) of the
particles in the unperturbed disk to the Eulerian spherical
coordinates ($r,\phi,\theta$) of the same particles in the perturbed
disk; $\bar\theta=\pi/2$ for the unperturbed disk. Thus, the map is
given by
\begin{eqnarray}\label{coo}
  r(t)=R(\bar r,\bar\phi;t), \ \ \phi(t)=P(\bar r,\bar\phi;t),\ \
  \theta(t)=T(\bar r,\bar\phi;t)\ ,
\end{eqnarray}
where $R$, $P$, and $T$ are functions we will parameterize below.

We require the mapping to be smooth and single-valued so that there
are no intersections of particle orbits, which would lead to shock
waves and invalidate our assumption that the disk remains cold and
thin.  In other words, the perturbed disk is given by a
diffeomorphism of the unperturbed disk. The number density
$n(r,\phi,\theta;t)$ of the perturbed disk follows from the
transformation of the disk area element $d\Sigma$. In the absence of
intersecting orbits, the number of particles $dN$ contained in each
surface element is constant: $dN=n_0(\bar r)d\bar
\Sigma=n(r,\phi,\theta;t)d\Sigma$. The surface number density of the
perturbed disk is then given by
\begin{eqnarray}\label{nn0}
n(r,\phi,\theta;t)=n_0(\bar r)\frac{d\bar \Sigma}{d\Sigma}
\end{eqnarray}
Here $d\bar \Sigma$ is an area element in the unperturbed disk. The
requirement that no intersections occur translates as a condition on
$d\Sigma$: $d\Sigma\neq 0$.

If the disk perturbations were coplanar displacements in flat space,
the ratio $d\Sigma/d\bar\Sigma$ would be a simple Jacobian
determinant.  However, the displacements are (in general)
three-dimensional and we allow for space curvature.  Thus a more
careful calculation of the area element is needed.  We do this using
an orthonormal basis constructed on slices of constant coordinate
time ($dt=0$).  The line element on each time slice is
$dl^2=dx^idx^jg_{ij}=dx^{\hat i}dx^{\hat j}\delta_{\hat i\hat j}$,
where the hatted indices give the components of a vector in the
orthonormal basis and Latin indices are summed over the spatial
components (with positive signature).  Here, $g_{ij}$ are the
spatial components of the full spacetime metric.

Orthonormal and coordinate components of three-vectors are related
by a triad $e^{\hat i}_{\ j}$: $dx^{\hat i}=e^{\hat i}_{\ j}dx^j$.
From the invariance of $dl^2$, we find that $e^{\hat i}_{\ k}e^{\hat
j}_{\ l}\delta_{\hat i \hat j}=g_{kl}$. For a diagonal spatial
metric $g_{ij}$ (i.e., an orthogonal coordinate basis) we can orient
the triad so that its only non-zero components are $e^{\hat i}_{\
i}=\sqrt{g_{ii}}$ (no summation). Throughout the rest of the paper,
we will assume that the spatial metric is diagonal (as it is for the
Kerr metric in Boyer-Lindquist coordinates).

Let us call the three spatial coordinates $(r,\phi,\theta)$ where
$\theta=\pi/2$ is the usual equatorial surface. The components of
the infinitesimal displacement vector \boldmath$v$\unboldmath$=d$\boldmath$x$\unboldmath, in the
right-handed coordinate basis
$(\partial_r,\partial_\phi,-\partial_\theta)$ are
$v^i=(dr,d\phi,-d\theta)$. Using the triad, the vector components in
the orthonormal basis\footnote{Note that we define
$\hat{\mbox{\boldmath{$\theta$}}}$ to point in the $-\partial_\theta$ direction}
$(\hat{\mbox{\boldmath{$r$}}},\hat{\mbox{\boldmath{$\phi$}}},\hat{ \mbox{\boldmath{$\theta$}}})$ are therefore
given by
\begin{eqnarray}\label{unit}
  v^{\hat i}=(Adr,Bd\phi,-Cd\theta)\ ,\ \
  A\equiv\sqrt{g_{rr}}\ ,\ \ B\equiv \sqrt{g_{\phi\phi}}\ ,\ \
  C\equiv \sqrt{g_{\theta\theta}}\ .
\end{eqnarray}
The factors $(A,B,C)$ in general can be functions of
$(r,\phi,\theta)$.  Every vector is associated with a point
$(r,\phi,\theta)$ in the perturbed disk and $(\bar r,\bar\phi)$ in
the unperturbed disk.

Let us construct two displacement vectors $\bar{\mbox{\boldmath{$X$}}}$ and
$\bar{\mbox{\boldmath{$Y$}}}$ whose origin is the point $(\bar
r,\bar\phi,\bar\theta)$ in the plane of the unperturbed disk as
follows,
\begin{eqnarray}
  \bar{\mbox{\boldmath{$X$}}}=\bar A d\bar r\hat{\mbox{\boldmath{$r$}}}, \ \
  \bar{\mbox{\boldmath{$Y$}}}=\bar B d\bar \phi\hat{ \mbox{\boldmath{$\phi$}}}\ ,
\end{eqnarray}
where $\bar A\equiv A(\bar r,\bar \phi,\bar \theta)$, and
analogously for $\bar B$. Under the diffeomorphism (\ref{coo}), we
find that the vectors are mapped to
\begin{eqnarray}
  \mbox{\boldmath{$X$}}=A\left(\frac{\partial R}{\partial \bar r}
    \right)_{\bar \phi}d\bar r\hat{\mbox{\boldmath{$r$}}}+B \left(
    \frac{\partial P}{\partial\bar r}\right)_{\bar \phi}
    d\bar r\hat{\mbox{\boldmath{$\phi$}}}-C \left(\frac{\partial T}
    {\partial \bar r}\right)_{\bar \phi}d\bar r
    \hat{\mbox{\boldmath{$\theta$}}} \nonumber\\
  \mbox{\boldmath{$Y$}}=A\left(\frac{\partial R}{\partial \bar\phi}
    \right)_{\bar r}d\bar\phi\hat{\mbox{\boldmath{$r$}}}+ B \left(
    \frac{\partial P}{\partial \bar\phi}\right)_{\bar r}
    d\bar\phi\hat{\mbox{\boldmath{$\phi$}}}-C \left(\frac{\partial T}
    {\partial \bar\phi}\right)_{\bar r}d\bar\phi\hat{\mbox{\boldmath{$\theta$}}}\ .
\end{eqnarray}
Note that the origin of the vectors is changed under the map, and
therefore $A,B,C$ are no longer barred.

The area of the parallelogram in the unperturbed disk formed by
$\bar{\mbox{\boldmath{$X$}}}$ and $\bar{\mbox{\boldmath{$Y$}}}$ is given by
$d\bar\Sigma=\left|\bar{\mbox{\boldmath{$X$}}}\times\bar{\mbox{\boldmath{$Y$}}}\right|=\bar A\bar
Bd\bar rd\bar \phi$. This parallelogram is mapped under (\ref{coo})
to a parallelogram with area $d\Sigma=|\mbox{\boldmath{$X$}}\times\mbox{\boldmath{$Y$}}|$. Note
the vector $\mbox{\boldmath{$X$}}\times\mbox{\boldmath{$Y$}}$ is normal to the perturbed disk.
Thus, we have
\begin{eqnarray}\label{dA}
  \frac{\mbox{\boldmath{$X$}}\times\mbox{\boldmath{$Y$}}}{|\bar{\mbox{\boldmath{$X$}}}\times\bar{\mbox{\boldmath{$Y$}}}|}
  &=&\frac{1}{\bar A\bar Bd\bar rd\bar \phi}
  \left|
\begin{array}{ccc} 
\hat{\mbox{\boldmath{$r$}}} & \hat{\mbox{\boldmath{$\phi$}}} & \hat{\mbox{\boldmath{$\theta$}}}\\
    A(\partial R/\partial \bar r)d\bar r &
    B(\partial P/\partial \bar r)d\bar r &
    -C(\partial T/\partial \bar r)d\bar r\\
    A(\partial R/\partial \bar \phi)d\bar\phi &
    B(\partial P/\partial \bar \phi)d\bar\phi &
    -C(\partial T/\partial \bar \phi)d\bar\phi
  \end{array}
\right|\nonumber\\
  &=&-\frac{BC}{\bar A\bar B} J_{PT}\hat{\mbox{\boldmath{$r$}}}+\frac{AC}{\bar A\bar B}
    J_{RT}\hat{\mbox{\boldmath{$\phi$}}}+\frac{AB}{\bar A\bar B}J_{RP}\hat{\mbox{\boldmath{$\theta$}}}\ ,
\end{eqnarray}
and therefore, from (\ref{nn0}),
\begin{eqnarray}\label{R}
  \left(\frac{n_0}{n}\right)^2=\left(\frac{d\Sigma}{d\bar \Sigma}\right)^2
  =\left(\frac{BC J_{PT}}{\bar A\bar B}\right)^2+\left(\frac{AC J_{RT}}
  {\bar A\bar B}\right)^2+\left(\frac{ABJ_{RP}}{\bar A\bar B}\right)^2\ ,
\end{eqnarray}
where $J_{MN}$ is the Jacobian of any functions $M(\bar r,\bar
\phi)$ and $N(\bar r,\bar \phi)$.

In order to avoid caustics, we must impose that $(n_0/n)^2\neq 0$ in
the disk at each moment in time.  For the spacetimes of interest,
the triad factors $(A,B,C)$ are nonzero in the disk, implying that
at least one of the three differential inequalities $J_{MN}\neq 0$
must always be satisfied at each point of the disk. In order to
systematically solve these inequalities, we will work in the
epicyclic approximation. Throughout the rest of the paper we will
assume that $A$, $B$ and $C$ do not vanish in the plane of the disk.

\section{Epicyclic Approximation}
For a generic potential the trajectories of test particles will not
form closed non-circular orbits at all radii. Since we want to avoid
intersections of the particle orbits, any deformation we make to the
orbits in the disk must be a perturbation to a circular orbit. In
this case we can use the epicyclic approximation, in which radial
and vertical perturbations are harmonic oscillations with coordinate
frequencies $\omega_r$ and $\omega_\theta$, respectively. In this
approximation, to first order in the perturbations $\epsilon$ and
$\gamma$, the map (\ref{coo}) can be written as
\begin{mathletters}\label{epic}
\begin{eqnarray}
  r(t)&=&R(\bar r,\bar\phi;t)=\bar r-\epsilon(\bar r,\Phi)
    \cos[\omega_r(\bar r)t+\chi(\bar r,\Phi)]\ ,
  \label{epicr}\\
  \phi(t)&=&P(\bar r,\bar\phi;t)=\bar \phi+\phi_0(\bar r,\bar\phi)
    +\omega_\phi(\bar r)t+\beta(\bar r)\epsilon(\bar r,\Phi)
    \sin[\omega_r(\bar r)t+\chi(\bar r,\Phi)]\ ,
  \label{epicphi}\\
  \theta(t)&=&T(\bar r,\bar\phi;t)=\bar\theta-\gamma(\bar
    r,\Phi)\cos[\omega_\theta(\bar r)t+\zeta(\bar
    r,\Phi)]\ .\label{epictheta}
\end{eqnarray}
\end{mathletters}
The horizontal epicycles have amplitude $\epsilon$ and phase $\chi$;
the vertical epicycles have amplitude $\gamma$ and phase $\zeta$; in
addition the azimuthal position is perturbed by $\phi_0$. The
azimuth $\phi$ does not increase monotonically but undergoes a
sinusoidal variation related to the radial oscillation through the
conservation of angular momentum.  In the Newtonian limit,
$r^2\dot\phi=\hbox{constant}$ implies
\begin{eqnarray}
  \beta(\bar r)=\frac{2\omega_\phi(\bar r)}{\bar r\omega_r(\bar r)}\ .
\end{eqnarray}
For the Kerr metric in Boyer-Lindquist coordinates, conservation of
angular momentum gives
\begin{eqnarray}
  \beta(\bar r)=\frac{2\bar r\omega_\phi(\bar r)}{\omega_r(\bar r)}
  \frac{[1+\omega_\phi(\bar r)(a-3\sqrt{GM\bar r})]}{\bar r^2-2GM\bar r+a^2}\ ,
\end{eqnarray}
where $a$ is the black hole spin with $a>0$ for prograde orbits and
$a<0$ for retrograde orbits.

It's worth noting that in writing equations (\ref{epic}) we are
neglecting time-independent perturbations that arise from deforming
one circular orbit into another.  We could account for them simply
by modifying the unperturbed disk.  In our linearized treatment such
perturbations are uninteresting because they cause no time
variability.

In equations (\ref{epic}) we used the variable $\Phi(\bar r,\bar
\phi)\equiv \phi(t=0)=\bar\phi+\phi_0(\bar r,\bar\phi)+\beta(\bar
r)\epsilon[\bar r,\Phi(\bar r,\bar \phi)]\times$ $\sin\chi[\bar
r,\Phi(\bar r,\bar\phi)]$, which can be solved iteratively for
$\Phi$, but such a solution is not necessary for our calculation.
This expression can be rewritten using (\ref{epicphi}) as
\begin{equation}\label{Phit}
  \Phi(\bar r,\phi;t)=\phi-\omega_\phi t-\beta\epsilon
  [\sin(\omega_rt+\chi)-\sin\chi]\ .
\end{equation}
The variables $(\bar r,\Phi)$ give the original coordinates of
particles in the perturbed disk. The choice of functional dependence
of the perturbation functions on $\Phi$ will allow us to write
$r(t)$ and $\theta(t)$ as functions of the mixed coordinates $(\bar
r,\phi)$ which will give the correct caustic structure provided that
a certain initial condition is satisfied, as we will show below.

The Eulerian coordinates $r$ and $\theta$ at fixed $t$ must be
periodic in $\phi$. Therefore, using (\ref{epic}) we obtain
\begin{mathletters}\label{periodic}
\begin{equation}
  \epsilon(\bar r ,\Phi+2\pi n)=\epsilon(\bar r ,\Phi), \ \
  \gamma(\bar r ,\Phi+2\pi n)=\gamma(\bar r ,\Phi)\ ,
  \label{period}
\end{equation}
\begin{equation}
  \chi(\bar r ,\Phi+2\pi n)=\chi(\bar r ,\Phi)+2\pi q(n), \ \
  \zeta(\bar r ,\Phi+2\pi n)=\zeta(\bar r ,\Phi)+2\pi q(n)\ ,
  \label{nperiod}
\end{equation}
\end{mathletters}
for any integer $n$. The function $q(n)$ appearing in the two
places above need not be one and the same, but we used the same
symbol to economize notation. Since the perturbation functions must
be smooth, this implies that $q(n)$ depends only on $n$, and not on
$\bar r$ or $t$ for instance.

In the epicyclic approximation (\ref{epic}), for any two functions
$M(\bar r,\Phi)$ and $N(\bar r,\Phi)$ we obtain
\begin{eqnarray}\label{jacfinal}
  J_{MN}&=&\left(\frac{\partial M}{\partial \bar r}\right)_{\bar\phi}
  \left(\frac{\partial N}{\partial \bar \phi}\right)_{\bar r}
  -\left(\frac{\partial N}{\partial \bar r}\right)_{\bar \phi}
  \left(\frac{\partial M}{\partial \bar \phi}\right)_{\bar r}
  \nonumber\\
  &=&\left(\frac{\partial \Phi}{\partial \bar \phi}\right)_{\bar r}
  \left[\left(\frac{\partial M}{\partial \bar r}\right)_{\Phi}
  \left(\frac{\partial N}{\partial \Phi}\right)_{\bar r}
  -\left(\frac{\partial N}{\partial \bar r}\right)_{\Phi}
  \left(\frac{\partial M}{\partial \Phi}\right)_{\bar r}\right]\ .
\end{eqnarray}
We do not want $J_{MN}$ to become zero, hence we require $(\partial
\Phi/\partial \bar \phi)_{\bar r}\neq 0$. This requires that the
initial conditions in the perturbed disk are such that the angle
$\Phi=\phi(t=0)$ is a function of $\bar \phi$, otherwise from
(\ref{epic}) we see that the map will no longer be one-to-one, which
in turn would lead to caustics in the initial conditions. Since this
term is time independent, once these initial conditions are
satisfied, the zeros of $J_{MN}$ and $J_{MN}/(\partial \Phi/\partial
\bar \phi)_{\bar r}$ will coincide. Thus, $J_{MN}\ne0$ is a
sufficient condition to justify our usage of $(\bar r,\Phi)$ to
label the particles in the disk.

In complete analogy as above, we find that the ratio of $J_{MN}$ to
the Jacobian of the two functions with $(\bar r,\phi)$ chosen as
independent variables is equal to $(\partial \Phi/\partial
\phi)_{\bar r,t}$.  From (\ref{Phit}), $(\partial \Phi/\partial
\phi)_{\bar r,t}-1$ is first order (or smaller) in $\beta\epsilon$
and $\beta(\partial\epsilon/\partial\Phi)_{\bar r}$. Below we will
show that for long-lived perturbations
$(\partial\epsilon/\partial\Phi)_{\bar r}=0$, and therefore the
caustic structure is preserved if we restrict ourselves to
small-amplitude perturbations ($\beta\epsilon\ll 1)$ and we choose
to label the particles with coordinates $(\bar r,\phi)$ or $(\bar
r,\Phi)$.

\section{Perturbations in the Disk Plane}

Having developed the necessary background for kinematic density
waves in possibly warped disks in flat or curved spacetime, we now
ask whether there exist perturbations that last many orbital times
before orbits intersect.  We will find in general that orbits always
intersect eventually in the absence of pressure or other
non-gravitational forces.  However, some patterns last many orbital
times. We begin the examination in this section with perturbations
that leave the disk flat.

\subsection{Long-lived Patterns and their Lifetimes}

Perturbations in the plane of the disk satisfy $T=\bar\theta=\pi/2$.
Thus, from (\ref{R}) we see that to avoid caustics we need
$J_{RP}\neq 0$. Let us write the Jacobian of the two functions
$[R(\bar r,\bar\phi;t),P(\bar r,\bar\phi;t)]$ in terms of the
Jacobian of the functions $[R(\bar r,\Phi;t),P(\bar r,\Phi;t)]$
defined by composition with $\bar\phi(\bar r,\Phi)$.  With $\Phi$ as
the angular variable the calculation simplifies somewhat as the
unknown function $\phi_0(\bar r,\bar\phi)$ is eliminated.

Using (\ref{epicr}), (\ref{epicphi}) and (\ref{jacfinal}) we can
obtain $J_{RP}/(\partial \Phi/\partial \bar \phi)_{\bar r}$. The
Jacobian has many terms, and to simplify things note that from
(\ref{periodic}), the derivatives of all perturbation functions with
respect to $\bar r$ and $\Phi$ are periodic in $\Phi$. Also, we have
to drop all terms proportional to $\epsilon^2$, since our
approximation does not conserve angular momentum at this order.
However, this does not imply that we can drop terms such as
$\epsilon(\partial\epsilon/\partial\bar r)$ and
$\epsilon(\partial\epsilon/\partial\Phi)$ since the derivatives need
not be small.

At $t=0$ the Jacobian evaluates to
\begin{equation}\label{jrp0}
  \frac{J_{RP}(t=0)}{(\partial \Phi/\partial \bar \phi)_{\bar r}}
  =1+\epsilon\frac{\partial\chi}{\partial\bar r}\sin\chi
    -\frac{\partial\epsilon}{\partial\bar r}\cos\chi\ .
\end{equation}
All partial derivatives with respect to $\bar r$ are taken holding
$\Phi$ constant.  In order to avoid caustics in the perturbed disk
initially we require $J_{RP}\ne0$ everywhere which will be satisfied
if
\begin{equation}\label{rders}
  \left(\frac{\partial\epsilon}{\partial\bar r}\right)^2
  +\left(\epsilon\frac{\partial\chi}{\partial\bar r}\right)^2
  <1\ .
\end{equation}
Similarly, to avoid caustics forming within an orbit time it is
sufficient to require
\begin{equation}\label{Phiders}
  \left(\beta\frac{\partial\epsilon}{\partial\Phi}\right)^2
  +\left(\beta\epsilon\frac{\partial\chi}{\partial\Phi}
  \right)^2\lesssim1\ .
\end{equation}
If we assume that $\beta\epsilon\sim\epsilon/\bar r\ll1$ and the
inequalities (\ref{rders}) and (\ref{Phiders}) are strongly
satisfied, then at late times the Jacobian becomes
\begin{equation}\label{jrp}
  \frac{J_{RP}(t\to\infty)}{(\partial \Phi/\partial \bar \phi)_{\bar r}}
  =1+\omega'_\phi t\left[\epsilon\left(\mu-\frac{\partial\chi}{\partial
  \Phi}\right)\sin(\omega_rt+\chi)+\frac{\partial\epsilon}
  {\partial\Phi}\cos(\omega_rt+\chi)\right]\ ,
\end{equation}
where a prime denotes a derivative with respect to $\bar r$, e.g.
$\omega'_\phi=d\omega_\phi/d\bar r$, and
\begin{equation}\label{mudef}
  \mu(\bar r)\equiv\frac{\omega'_r}{\omega'_\phi}\ .
\end{equation}

To avoid intersection of orbits we require $J_{RP}>0$.  At late
times this requires
\begin{equation}\label{nocaus}
  \frac{\partial\chi}{\partial\Phi}\approx\mu(\bar r)\ ,\ \
  \frac{\partial\epsilon}{\partial\Phi}\approx0\ .
\end{equation}
These conditions may be integrated to give $\chi\approx\mu(\bar r)
\Phi+\eta(\bar r)$ and $\epsilon\approx\epsilon(\bar r)$ where
$\eta$ is an integration constant.  From (\ref{nperiod}), the first
condition implies $\mu=m$ is an integer.  Thus, to maximize the
chance of having long-lasting waves we restrict perturbations to
have the following form:
\begin{equation}\label{pertr}
  \epsilon=\epsilon(\bar r)\ ,\ \ \chi=m\Phi+\eta(\bar r)\ .
\end{equation}
The two functions $\epsilon(\bar r)$ and $\eta(\bar r)$ specify the
planar perturbations.

From (\ref{pertr}) we see that the first of conditions
(\ref{nocaus}) cannot be satisfied everywhere in the disk because
$\mu$ is a function of $\bar r$.  Thus, intersections will
inevitably occur: it is impossible to have an infinite-lived
kinematic density wave except for special power-law potentials like
Kepler and the simple harmonic oscillator.  However, the timescale
for caustic formation can be made arbitrarily large if one restricts
perturbations to small neighborhoods around the discrete radii $r_m$
such that $\mu(r_m)=m$.  Thus we consider a perturbation
$\epsilon(\bar r)$ which is negligible everywhere except over a
range $\delta r$ about $r_m$.  From (\ref{jrp}) and (\ref{pertr}),
the perturbation lifetime until caustic formation is
\begin{equation}\label{Tcaus}
  \tau_c=\frac{1}{\epsilon\omega'_\phi(\mu-m)}\approx
    \frac{1}{\epsilon\omega'_\phi\mu'\delta r}\ .
\end{equation}
In linear perturbation theory, $\delta r$ should be interpreted as
$\delta\bar r$, i.e.\ the initial displacement of the unperturbed
disk.  However, it is more accurate to evaluate the orbital
frequencies at the displaced position, in which case one can regard
$\delta r$ as the maximum of the amplitude $\epsilon_{\rm max}$ and
the width $\delta\bar r$ of the perturbation field $\epsilon(r)$
(which might be, for example, a Gaussian centered at $r_m$).

Thus, we have found relatively long-lived perturbations around
discrete radii $r_m$ satisfying $\mu(r_m)=m=$ integer. Combining
(\ref{epic}), (\ref{Phit}), and (\ref{pertr}) we find that to first
order in $\epsilon$ the perturbations have the form:
\begin{equation}\label{rsol}
  r=\bar r-\epsilon(\bar r)\cos[m\phi+\omega_m(\bar r)t+\eta(\bar r)], \ \
  \omega_m(\bar r)\equiv\omega_r(\bar r)-m\omega_\phi(\bar r)\ .
\end{equation}
This perturbation represents an $|m|$-armed spiral and $\omega_m$ is
called the modulation speed.\footnote{This is the angular speed of
light curve oscillations, as opposed to the lower angular speed for
the pattern to rotate a full $2\pi$.  The corresponding pattern
speed is $\omega_p=-\omega_m/m=\omega_\phi-\omega_r/m$.
\cite{Binney1987} define $\omega_p=\omega_\phi-(n/m)\omega_r$ where
$n$ and $m$ are integers. This is exactly the result of \cite{Amin2006}. For $n>1$ the orbits self-intersect so we
do not consider this case.} The modulation frequency of the spiral
is $f_m=\omega_m/2\pi$, which is the frequency for an arm-to-arm
displacement. Note that $\omega'_m=0$ at $r=r_m$; the modulation
speed has an extremum at the radii of long-lived perturbations.

In order to produce observable modulation, the pattern must not only
survive against intersections, it must also not be too tightly
wound.  If, for fixed $\phi$ and $t$, $r-\bar r$ oscillates rapidly
in the radial direction, the light curve will show little
modulation.  Significant modulation can persist only for a
winding-up timescale defined by the condition that the argument of
the cosine in (\ref{rsol}) change by less than $2\pi$ as $r$ changes
for fixed $\phi$ and $t$.  At arbitrary $\bar r$, this gives the
condition $|t[\omega_m(\bar r+\delta\bar r)-\omega_m(\bar
r)]+\eta'\delta\bar r|<2\pi$. At $\bar r=r_m$, $\omega_m'=0$ and the
lifetime against winding is (if $\eta'$ is neglected)
\begin{equation}\label{Twind}
  \tau_w=\frac{4\pi}{\omega''_m(\delta\bar r)^2}\sim
  \frac{2\pi}{\omega_\phi'\mu'(\delta r)^2}\ .
\end{equation}
The long lifetime has a simple interpretation.  If the pattern speed
$\omega_p=-\omega_m/m$ is constant over a range of radii, a density
wave located over that range will remain constant in the frame
rotating with angular velocity $\omega_p$, hence will avoid winding
up.  Note that if $\delta r\sim\epsilon$ (larger $\epsilon$ is
forbidden by the requirement $\epsilon'<1$), the timescales for
winding up and caustic formation are comparable; in general winding
up precedes caustic formation.  Had we retained nonzero $\eta'$, the
winding up would have been faster on one side of $r=r_m$ and slower
on the other.

\subsection{Spiral Arm Direction}

In order to see whether the spiral arms are leading or trailing, we
need to find the sign of $(\partial \phi/\partial
r)_{\mathrm{spiral\ arms}}$ evaluated along the spiral arms. Thus,
we first find the initial spiral arm positions.  These follow from
the extrema of $J_{RP}(\bar r,\Phi;t=0)$.  From (\ref{jrp0}) and
(\ref{pertr}), setting $\eta=0$ to maximize the perturbation
lifetime, we obtain
\begin{equation}\label{n0}
  \frac{n}{n_0}\approx\frac{1}{1-\epsilon'\cos(m\Phi)}\ .
\end{equation}
Setting the gradient to zero gives conditions for extrema:
$\epsilon''\cos(m\Phi)=0$ and $m\epsilon'\sin(m\Phi)=0$. For generic
$\epsilon$ and $m\neq 0$ this corresponds to the loci: (1)
$\epsilon'=0$ and $\Phi=(N+1/2)\pi/m$; and (2) $\epsilon''=0$ and
$\Phi=N\pi/m$; where $N$ is an integer. From the second partial
derivatives we find that the first locus corresponds to saddle
points in the density, while the second corresponds to overdensities
if $(-1)^N\epsilon'>0$ and $-(-1)^N\epsilon'''>0$ at these points.
If the signs are reversed, then we would have underdensities; and if
the signs are mixed then we have a saddle. An example of this result
is given in the first snapshot of Figure \ref{fig:disk} for the
$m=-2$ mode in the Kerr metric. There we have a gaussian
$\epsilon(\bar r)$ for which we obtain four minima and four maxima
in the density. If at each radius we maximize $n/n_0$ with respect
to $\Phi$, we find that initially the spiral arms are radial
segments lying at $\Phi=N\pi/m$. Thus, there are $2|m|$
overdensities for the mode with azimuthal number $m$.

The pattern makes a $2\pi$ revolution with a period $2\pi
m/\omega_m$, which implies that $\phi+\omega_m t /m =N\pi/m$ along
the spiral arms. Since we need only the sign of $(\partial
\phi/\partial r)_{\mathrm{spiral\ arms}}$, we can work to zeroth
order in $\epsilon$. The sense of winding up of each spiral arm is
then given by
\begin{eqnarray}\label{leadtrail}
 \left(\frac{\partial \phi}{\partial \bar r}\right)_{\phi=
   (N\pi-\omega_m t)/m}=\omega_p't\equiv-\frac {\omega_m't}{m}
   =-\frac{\omega_\phi't}{m}(\mu-m)\ .
\end{eqnarray}
Thus, assuming that $\omega_m(\bar r)$ is monotonic across the
perturbation, we can conclude that overdensities which lie inside
$r_m$ have opposite sense of winding than those lying outside.  This
is consistent with our earlier result that the pattern speed changes
sign across the radii $r_m$. This behavior is exemplified Figure
\ref{fig:disk} for the $m=-2$ mode in the Kerr metric.  Leading
waves correspond to a pattern speed increasing with radius, or
$\omega_m'<0$.

\subsection{Spherical potentials in the Newtonian limit}

Applying the results obtained above requires finding integer
solutions of $\mu(\bar r)=m$.  In the Newtonian limit, with
gravitational potential $V(\bar r)$, the effective potential for
radial motion is $V_{\mathrm{eff}}(\bar r)=V(\bar
r)+L^2/2\bar{r}^2$. For a unit mass test particle we can then write
\begin{eqnarray}
  L&=&\omega_\phi \bar{r}^2, \ \
  \omega_\phi=\sqrt{\frac{1}{\bar r}V'(\bar r)},\ \
  \omega_r=\sqrt{\left(V_{\mathrm{eff}}''
    (\bar r)\right)_{L=\mathrm{const}}}
    =\sqrt{\frac{3}{\bar r}V'(\bar r)+V''(\bar r)}\\
  \mu&=&\frac{\omega_\phi}{\omega_r}\left[3-\frac{\bar{r}^2V'''
    (\bar r)}{V'(\bar r)-\bar rV''(\bar r)}\right]\ ,
\end{eqnarray}
in the Newtonian limit. Note that the derivatives in
$\mu=\omega_r'/\omega_\phi'$ are taken across particle orbits and
thus are not evaluated at $L=$const.

As  an example, we consider a set of classical potentials used in
the literature. Surprisingly, we find that they are divided into
three classes:

\begin{enumerate}
\item Potentials which do not have integer solutions for $\mu$. These
include: power law potentials $V(r)\propto -1/r^n$ with $2>n> 1$,
the Hernquist potential $V(r)=-a/(r+b)$, the Jaffe potential
$V(r)\propto \ln(r+a)$, and the NFW potential
$V(r)=-(a/r)\ln(1+r/b)$, where $a$ and $b$ are constants.  These
potentials admit no long-lived kinematic density waves.

\item Potentials with one solution for $\mu=2$ (and another one for
$\mu=3$ at $r=0$): the potential
$V(r)=-(a/r)\ln[r/b+\sqrt{q+(r/b)^2}]$, the logarithmic halo
potential $V(r)\propto \ln(r^2+a^2)$, and the Plummer potential
$V(r)\propto -1/\sqrt{r^2+a^2}$.

\item Potentials with non-positive integer solutions for $\mu$:
$V(r)=-b[1+(a/r)^2]/r$, $V(r)\propto 1-\exp(a/r)$, and the effective
potential of the Kerr metric, which is discussed below.
\end{enumerate}

\section{Vertical and Three-dimensional Perturbations}

Next we consider perturbations that warp the disk out of its plane,
starting with the case $\epsilon=0$ but $\gamma\ne0$. The discussion
of vertical perturbations follows closely the earlier presentation.
The relevant Jacobians are given by:
\begin{mathletters}
\begin{eqnarray}
  \frac{J_{RT}}{(\partial \Phi/\partial \bar \phi)_{\bar r}}
  &=&\frac{\partial\gamma}{\partial\Phi}\cos\left(\omega_\theta t
    +\zeta\right)-\gamma\frac{\partial\zeta}{\partial\Phi}
    \sin\left(\omega_\theta t+\zeta\right), \label{jrt}\\
  \frac{J_{RP}}{(\partial \Phi/\partial \bar \phi)_{\bar r}}
  &=&1\label{Jrp}\\
  \frac{J_{PT}}{(\partial \Phi/\partial \bar \phi)_{\bar r}}
  &=&\mathcal{E} \cos\left(\omega_\theta t+\zeta\right)
  +\mathcal{F}\sin\left(\omega_\theta t+\zeta\right)\label{jpt}\ ,
    \label{Jpt}
\end{eqnarray}
\end{mathletters}
where
\begin{equation}\label{efnudef}
  \mathcal{E}\equiv \omega_\phi't\frac{\partial\gamma}{\partial\Phi}
    -\frac{\partial\gamma}{\partial\bar r}, \ \
  \gamma^{-1}\mathcal{F}\equiv\omega_\phi' t\left(\nu-
    \frac{\partial\zeta}{\partial\Phi}\right)
    +\frac{\partial\zeta}{\partial\bar r}, \ \
  \nu\equiv\omega_\theta'/\omega_\phi'\ .
\end{equation}
For orbits in a spherical potential, $\omega_\theta=\omega_\phi$
implying $\nu=1$ for all $r$.  For a nonspherical potential,
$\nu\ne1$.

As long as we require that $(\partial \Phi/\partial \bar \phi)_{\bar
r}\neq 0$, we have $J_{RP}\neq 0$, implying that no caustics will
ever form since at least one contribution to (\ref{R}) is nonzero.
By comparing (\ref{jrp}) with (\ref{jpt}), we can expect to find a
spiral pattern for the vertical perturbations. Let us find and
maximize the winding-up timescale.  This occurs when the
oscillations of $J_{PT}$ are bounded in amplitude as $t\to\infty$.
By analogy with equations (\ref{nocaus}) and (\ref{pertr}) we obtain
\begin{equation}\label{pertv}
  \gamma=\gamma(\bar r)\ ,\ \ \zeta=n\Phi+\xi(\bar r)\ ,
\end{equation}
where $n$ is an integer. The two functions $\gamma(\bar r)$ and
$\xi(\bar r)$ specify the vertical perturbations.  Similarly to
equation (\ref{rsol}), we find
\begin{equation}\label{rsoln}
  \theta=\bar\theta-\gamma(\bar r)\cos[n\phi+\Omega_m(\bar r)t
    +\xi(\bar r)], \ \
  \Omega_n(\bar r)\equiv\omega_\theta(\bar r)-n\omega_\phi(\bar r)\ .
\end{equation}
As before, the pattern is an $|n|$-armed spiral whose modulation
speed has an extremum at the radii of long-lived perturbations:
$\Omega'_m=0$ at $r=r_n$.

The vertical winding-up timescale is defined by the condition that
the argument of the cosine in (\ref{rsoln}) changes by less than
$2\pi$ as $r$ changes for fixed $\phi$ and $t$.  By analogy with the
planar case, near the critical radii where $\nu(\bar r)=n$ we find
(for nonspherical potentials and $\xi'=0$)
\begin{equation}\label{Twn}
  \tau_w=\frac{4\pi}{\Omega''_m(\delta\bar r)^2}\sim
  \frac{2\pi}{\omega_\phi'\nu'(\delta r)^2}\ .
\end{equation}
A nonzero $\xi'$ would decrease $\tau_w$ on one side of $r_n$.  For
a spherical potential, $\Omega_m=0$ for $n=1$ corresponding to a
stationary warp.  Such a perturbation is long-lived but provides no
modulation for a light curve.  A weak nonspherical distortion of the
potential or a weak non-gravitational torque would produce a nonzero
pattern speed, hence modulation, and might still preserve a long
lifetime against winding up.

Note that the Jacobian $J_{RT}$ (\ref{jrt}) is well-behaved and does
not lead to further constraints.

Now let us consider general orbit perturbations with $\gamma\neq
0\neq \epsilon$. The first question that arises is whether we can
still use the previous results. To see that this is indeed the case,
let us find the relevant three Jacobians. The Jacobian $J_{PT}$ is
exactly the same as in (\ref{jpt}), and the Jacobian $J_{RP}$ is the
same as the one in (\ref{jrp}).  The Jacobian $J_{RT}$ is quite
complicated in general, but it simplifies in the limit of large
times:
\begin{equation}\label{Jrt}
  \frac{J_{RT}(t\to\infty)}{(\partial\Phi/\partial\bar\phi)_{\bar r}}
  =\epsilon\gamma\omega'_\phi t(\mu n-\nu m)
    \sin(\omega_r t+\chi)\sin(\omega_\theta t+\zeta)\ ,
\end{equation}
where $\mu=\omega_r'/\omega_\phi'$ and
$\nu=\omega_\theta'/\omega_\phi'$ as before.  The integers $m$ and
$n$ follow as before from periodicity conditions on $\chi$ and
$\zeta$.

The condition for caustics to form is that all three Jacobians
vanish at the same point.  In general this is impossible. However,
perturbations can become tightly wound, after which time they would
produce little modulation in a light curve. Maximizing the winding
up time leads to the epicyclic solutions (\ref{rsol}) and
(\ref{rsoln}) found before.

A necessary (but not sufficient) condition for the three-dimensional
wave to be long-lived (i.e., the inverse timescale is quadratic in
$\delta r/r$) is that the perturbations are centered around radii
$r_c$ which satisfy both $\mu(r_c)=m$ and $\nu(r_c)=n$. The
requirement that both $\mu(r_c)$ and $\nu(r_c)$ are integers is very
restrictive and is possible only for spherical potentials where
$\nu=1$ for all $r$, or other special cases.

\section{Accretion Disks Around Kerr Black Holes}

The Kerr metric in Boyer-Lindquist coordinates is stationary and
axisymmetric with orthogonal metric on constant time slices:
\begin{equation}\label{kerr}
  ds^2=-\alpha^2dt^2+\varpi^2(d\phi-\omega dt)^2
  +\frac{\rho^2}{\Delta}dr^2+\rho^2d\theta^2\ .
\end{equation}
In geometrized units $G=c=1$, for a black hole of mass $M$ and spin
$a$ (with $-M\le a\le M$ to allow for prograde or retrograde
orbits),
\begin{eqnarray}\label{boyerlin}
  \rho^2\equiv r^2+a^2\cos^2\theta\ ,\ \ \Delta\equiv r^2-2Mr+a^2\ ,\ \
  \alpha^2\equiv1-\frac{2Mr}{\rho^2}\ ,\nonumber\\
  \omega\equiv \frac{2Mra}{\rho^2\Delta+2Mr(a^2+r^2)}\ ,\ \
  \varpi^2\equiv\left[\frac{\rho^2\Delta+2Mr(a^2+r^2)}{\rho^2}\right]
    \sin^2\theta\ .\quad
\end{eqnarray}
Throughout this section, when we give frequencies or timescales in
SI units, those are given for a one solar mass black hole.  The
physical frequency is obtained by multiplying the value for the
solar-mass black hole by $M_\odot/M$.

On slices of constant coordinate time ($dt=0$), the functions $A$,
$B$, and $C$ in (\ref{unit}) can be read off from the metric:
\begin{eqnarray}
A=\frac{\rho}{\sqrt{\Delta}}, \ \ B=\varpi, \ \ C=\rho\ .
\end{eqnarray}
As desired, these quantities are nonzero outside the event horizon
except at the poles (a coordinate singularity).  The epicyclic
coordinate frequencies for circular orbits in the equatorial plane
of the Kerr metric are given by \citep{Perez1997}
\begin{mathletters}\label{kerr_freq}
\begin{equation}
  \omega_\phi=\frac{1}{r(r/M)^{1/2}+a}\ ,
\end{equation}
\begin{equation}
  \omega_r=\omega_\phi\left(1-6\frac{M}{r}+8a\sqrt{\frac{M}{r^3}}
    -3\frac{a^2}{r^2}\right)^{1/2}\ ,
\end{equation}
\begin{equation}
  \omega_\theta=\omega_\phi\left(1-4a\sqrt{\frac{M}{r^{3}}}
    +\frac{3a^2}{r^2}\right)^{1/2}\ .
\end{equation}
\end{mathletters}

\clearpage
\begin{figure}
\begin{center}
\includegraphics[width=10cm]{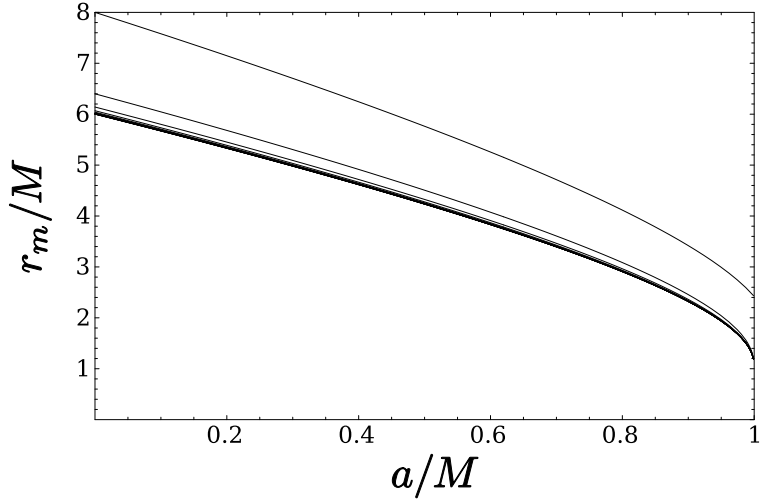}
\end{center}
\caption{The radii $r_m$ for planar density waves satisfying
$\mu(r_m)=m$, versus black hole spin for the Kerr solution. The
integer mode number $m$ is zero for the uppermost curve and then
decreases with decreasing $r_m$. As $m\to-\infty$, $r_m$ approaches
$r_\mathrm{ISCO}$, the innermost stable circular orbit.}
\label{fig:rm}
\end{figure}

Using (\ref{kerr_freq}) we can compute both $\mu$ and $\nu$. The
equation $\mu(r_m)=m$ has integer solutions for all $m\leq 0$, with
$m\to 1^-$ as $r\to\infty$, recovering the result for the Newtonian
Kepler potential. From (\ref{rsol}) we conclude that the spiral
patterns formed at $r_m$ move in the prograde direction (i.e., in
the same direction as the particle bulk flow) for $m\le0$. The radii
$r_m$ are plotted versus the black hole spin parameter, $a$, in
Figure \ref{fig:rm}. Notice that $r_m$ approaches the innermost
stable circular orbit (ISCO) for large $-m$. The properties of the
$m=0$ mode are qualitatively different from those of the modes with
$m\neq0$, and this can be seen in all plots that follow. The reason
is that when $m=0$ the pattern frequency is determined by the radial
oscillation frequency, while for the rest of the modes $\omega_m$ is
determined mainly by $\omega_\phi$.

\begin{figure}
\begin{center}
\includegraphics[width=10cm]{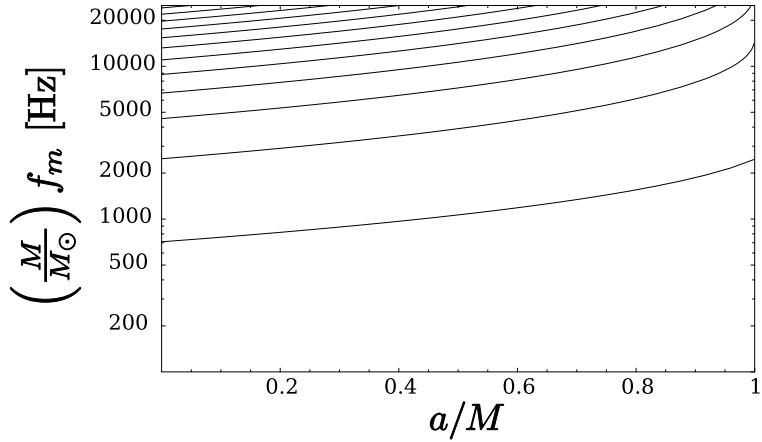}
\end{center}
\caption{The modulation frequency (the frequency of observed light
curve variations) for planar density waves of different $m$ in the
Kerr metric. The lowest frequency curve corresponds to $m=0$, with
the frequency increasing with $|m|$.  These frequencies appear to be
too large to account for observed black hole QPOs.} \label{fig:fm}
\end{figure}
\clearpage

Figure \ref{fig:fm} shows the modulation frequency
$f_m=\omega_m/2\pi$. The lowest mode $m=0$ has a frequency of
$f_m>700$ Hz, and the next mode has a frequency exceeding 2 kHz. One
can compare these frequencies with measured QPO frequencies for
well-studied black hole binaries \citep{Remillard2006}.  The lower
of the two commensurate frequencies are 300 Hz for GRO J1655-40, 184
Hz for XTE J1550-564, and 113 Hz for GRS 1915+105; for each case
there is a second frequency about 1.5 times higher.
\cite{Remillard2006} show that the lower of the two frequencies is
approximately 1860 $M_\odot/M$ Hz for these systems, too small to be
consistent with the modulation frequencies of Figure \ref{fig:fm}
for $|m|>0$. This is a shame, given that the frequency ratios for
$m=-2$ and $m=-3$ have nearly a 3:2 ratio over a wide range of black
hole spin. Thus, the simplest model of kinematic density waves fails
to account for the black hole high-frequency QPOs.

Our results for the pattern speed and $r_m$ of the $m$-th mode
exactly reproduce the frequency and radius of the same g-mode as
derived in the diskoseismology model of \cite{Perez1997} in the pressureless
limit (cf. our Figure \ref{fig:fm} and their
Figure 5). In the limit of large vertical mode number, the frequency difference between our and the diskoseismic result is proportional to the sound speed and the radial mode number, and is inversely proportional to the vertical mode number [as indicated in \cite{Perez1997}, equation (5.3)]. Since we consider only pressureless disks, we recover no analogs of p- and c-modes. However, our analysis of g-modes in pressureless disks is valid for generic potentials,
and it allows for the interpretation of the perturbations as
kinematic density spiral arms, as well as for the visualization of
the patterns that form in the disk. Thus, we are able to see that
the $m$-th mode has $2|m|$ overdensities (see below) resulting from
pure kinematics, which might double the expected frequency of the
X-ray modulations.  This result has been missed by the diskoseismic
models. The reason for that is that in the diskoseismic analysis, \cite{Perez1997} assume a time dependence of the Lagrangian perturbations of the form $e^{i\sigma t}$, where $\sigma$ is independent of $\bar r$. Comparing this with our equation (\ref{epicr}), we can read off the $\chi_{ds}$ for the diskoseismic models: $$\chi_{ds}=m\Phi+t(\sigma-\omega_r(\bar r))+\beta\epsilon\mathcal{O}(1).$$ Thus, the choice of time dependence in the diskoseismic analysis breaks the epicyclic approximation after a timescale of $$\tau_{ds}=\frac{1}{\delta r\mu\omega_\phi'}$$
which is $\sim(\delta r)^{-1}$ times shorter than the lifetime of the perturbations that we find. In the pressureless limit, nothing can break the validity of the epicyclic approximation. In the diskoseismic models this is reflected in the fact that the radial width of the perturbations vanishes in that limit. Therefore, our result generalizes the result of the diskoseismic models for pressureless disks by allowing for transients and maximizing their lifetimes.

\clearpage
\begin{figure}
\begin{center}
\includegraphics[width=8cm]{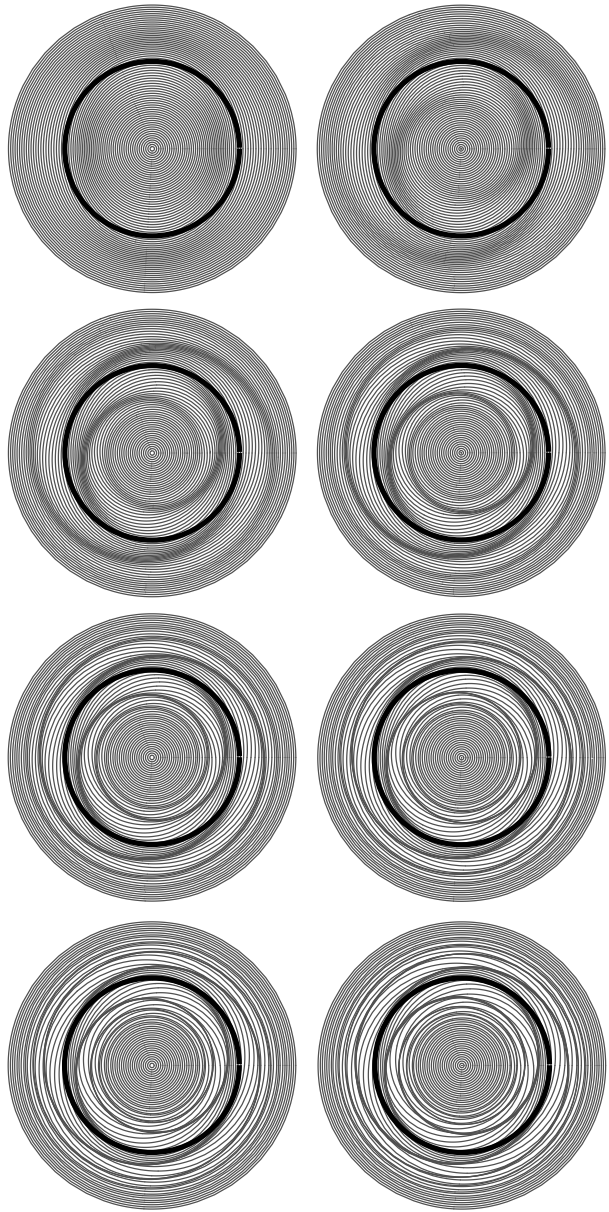}
\end{center}
\caption{Spiral-arm pattern caused by a perturbation with $m=-2$
around a Kerr black hole with a spin parameter $a=0.5M$.  Time
advances from left to right and top to bottom. The dark circle shows
the circle where $\mu(r)=-2$, i.e. $r_{-2}=4.33M$. The ISCO is
$r_\mathrm{ISCO}=4.23M$. The spiral pattern is formed by a
small-amplitude perturbation in the plane, as described in the text.
Since the perturbation is nonzero only in an extremely thin annulus
around the black hole, we have cut out the central portions of the
disk for clarity. Thus, the center of the polar plot does not
correspond to $r=0$, but to $r=r_{-2}-0.015M$, and the plot extends
to $r=r_{-2}+0.01M$.  The last two snapshots already have formed
caustics. Note that although $m=-2$, there are in fact 4 overdense
regions because the waves are leading inside $r_{-2}$ and trailing
outside $r_{-2}$.} \label{fig:disk}
\end{figure}
\clearpage

To illustrate the kinematics of the planar density waves, in Figure
\ref{fig:disk} we plot the $m=-2$ spiral pattern for several times.
The initial perturbation was chosen with $\eta(r)=0$ and
$\epsilon(r)=10^{-3}M\exp[-(r-r_{-2})^2/2\sigma^2]$, with $\delta
r\sim\sigma=0.003M$. The arm-to-arm displacement period is 0.87 ms.
The timestep between the snapshots is 9.3 s, and we have made it to
be an exact multiple of the pattern frequency. Therefore, the
pattern appears stationary, although it rotates counterclockwise.
The winding-up timescale is about 30 s, and the caustic-formation
timescale is about 1 minute. We ran similar calculations for a set
of perturbations with different $m$, $a$, $\epsilon$ and $\eta$, and
we found that our analytical estimates for the winding-up timescale
and the lifetime of the pattern are consistent with the numerical
results.

Figure \ref{fig:disk} verifies the analytic expectations of section
4.2. We see that for $\eta=0$ the spiral pattern for a mode with
mode number $m$ has $2|m|$ overdense regions --- $|m|$ trailing
waves on the outer edge of the perturbation and $|m|$ leading waves
on the inner edge in agreement with equation (\ref{leadtrail}).  If
these overdensities produce comparable modulation of the intensity,
then an observer would see a modulation at a frequency $2f_m$,
exacerbating the problem of matching kinematic density waves to
observed black hole QPOs.

If simple density waves cannot explain the observed QPOs, maybe a
beat frequency between an orbiting blob and density waves can do the
job. Consider a blob of gas orbiting the black hole near $r_m$ with
angular frequency $\omega_\phi(r_m)$.  It must be small and dense
enough not to be tidally sheared apart.  For $m\leq-1$ the spiral
arms will rotate faster than the blob around the disk and thus the
blob will be compressed every time the spiral waves pass over it.
This leads to a beat frequency in the emission from the blob given
by $(\omega_m-|m|\omega_\phi)/2\pi=\omega_r/2\pi\equiv f_r(r_m)$,
i.e.\ the radial epicyclic frequency. While the blob remains on one
side of $r_m$, its beats lead to a flickering frequency of $f_r$. If
the blob straddles $r_m$ hence is overtaken by both leading and
trailing density waves, the flickering frequency is $2f_r$.

\clearpage
\begin{figure}
\begin{center}
\plotone{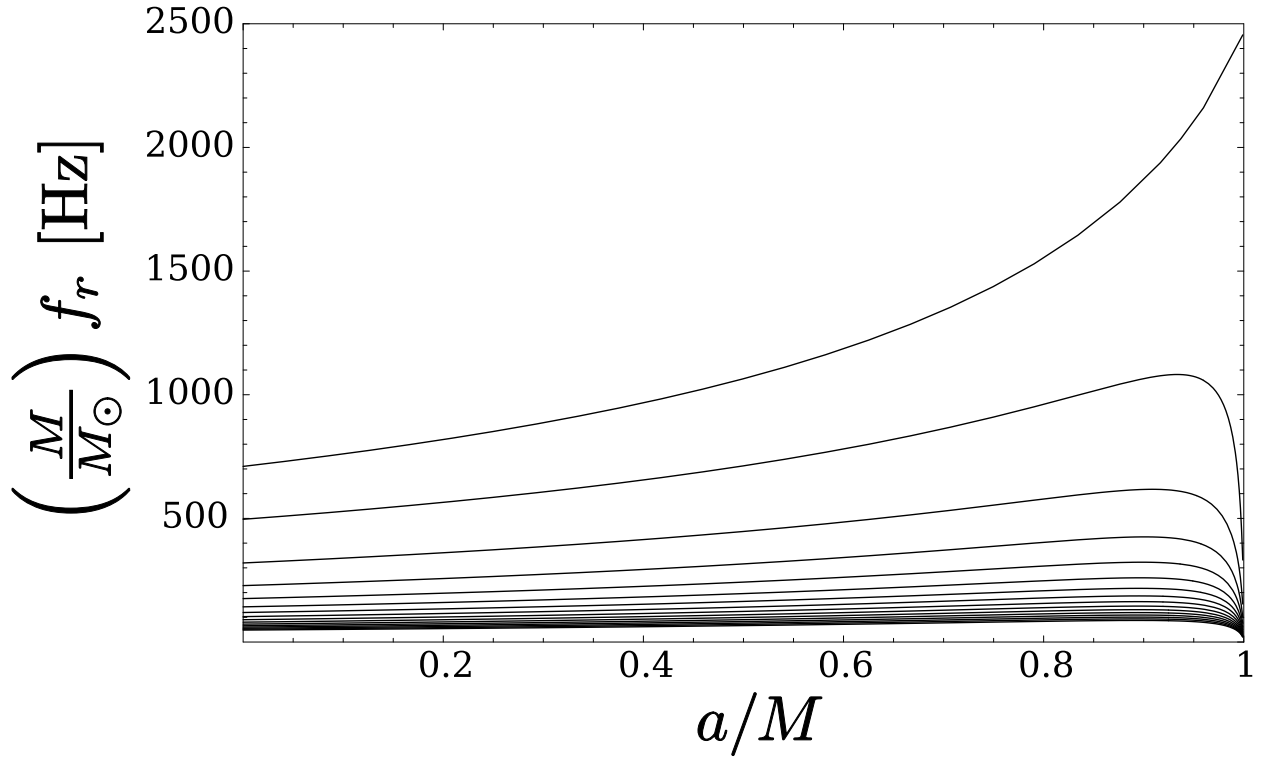}
\end{center}
\caption{The radial oscillation frequency at $r_m$ versus the black
hole spin parameter. The uppermost curve corresponds to $m=0$, with
$f_r$ decreasing as $m$ becomes more negative.  The beat frequency
between a blob in the disk and a kinematic density wave near $r_m$
is either $f_r$ or $2f_r$ (the frequency being doubled if the blob
straddles $r_m$).  These frequencies appear to be too small to
account for observed black hole QPOs.} \label{fig:fr}
\end{figure}
\clearpage

The radial frequency $f_r$ is plotted in Figure \ref{fig:fr}.  The
radial frequency in the Kerr metric is less than the azimuthal
frequency and goes to zero at the ISCO (corresponding to modes with
$m\to-\infty$).  This frequency is too low to account for
high-frequency QPOs, even if it is doubled.

\clearpage
\begin{figure}
\begin{center}
\plotone{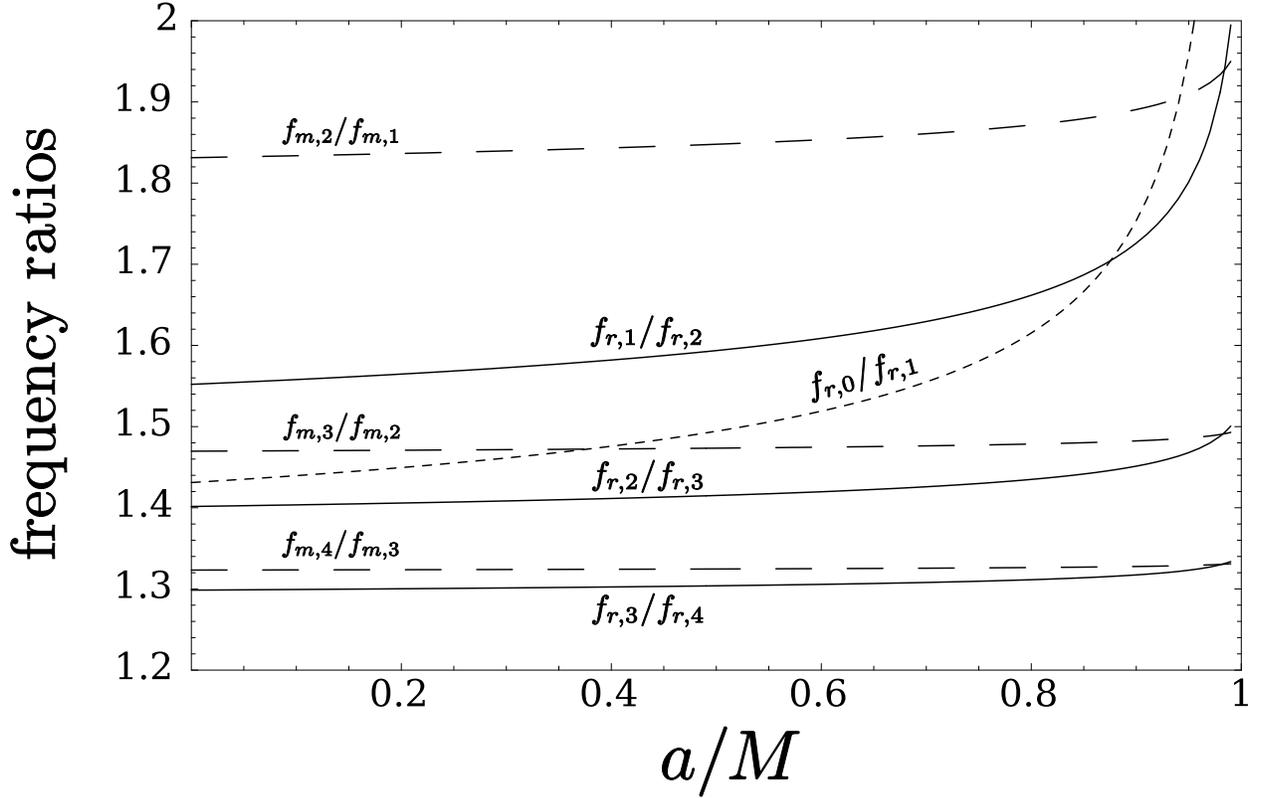}
\end{center}
\caption{Low-$m$ sets of frequencies having approximately a $3:2$
ratio.  The curves are labeled by the appropriate ratios of
$f_{m,|m|}$ for density waves or $f_{r,|m|}$ for blob-wave beat
frequencies (in all cases $|m|\le0$).  Ratios of $f_m/f_r$ exceed
2.0 except for the special case $f_{m,0}=f_{r,0}$.  Although the
pairs of modes with $m=-2$ and $m=-3$ have ratios close to 1.5 over
a wide range of spin parameter, the actual frequencies are
discrepant with observations.} \label{fig:frat}
\end{figure}
\clearpage

Maybe the black hole mass estimates are in error.  If so, the
predicted frequencies are wrong but frequency ratios are robust.
Thus in Figure \ref{fig:frat} we plot all frequency ratios of
low-$m$ modes that are close to 1.5.  Both the density wave
modulation frequencies $f_m$ and the radial (blob-wave beat)
frequencies $f_r$ are included. The only ratios that come close to
1.5 are the pairs with $m=-3$ and $m=-2$. However, the density wave
frequencies are too high (the lower frequency is at least 2500 Hz
for one solar mass) and the beat frequencies are too low (the lower
frequency is less than 420 Hz or 840 Hz if it is doubled) compared
with the observed range 1100 to 2000 Hz.  The nearest frequency
match occurs for spiral density waves with $m=-2$ for $a/M=0$, which
would require that the black hole masses be about 25\% larger than
the upper range of observational estimates.  For GRS 1915+105, for
example, the lower 113 Hz QPO could be explained by $f_{m,-2}$ if
the black hole mass were 22 $M_\odot$. The frequency ratio $168/113$
is close to $f_{m,-3}/f_{m,-2}=1.47$.  However, the spiral density
wave model also predicts modulation at frequencies twice as high
(from the combination of inner leading and outer trailing arms),
which has not been seen.

\clearpage
\begin{figure}
\begin{center}
\plotone{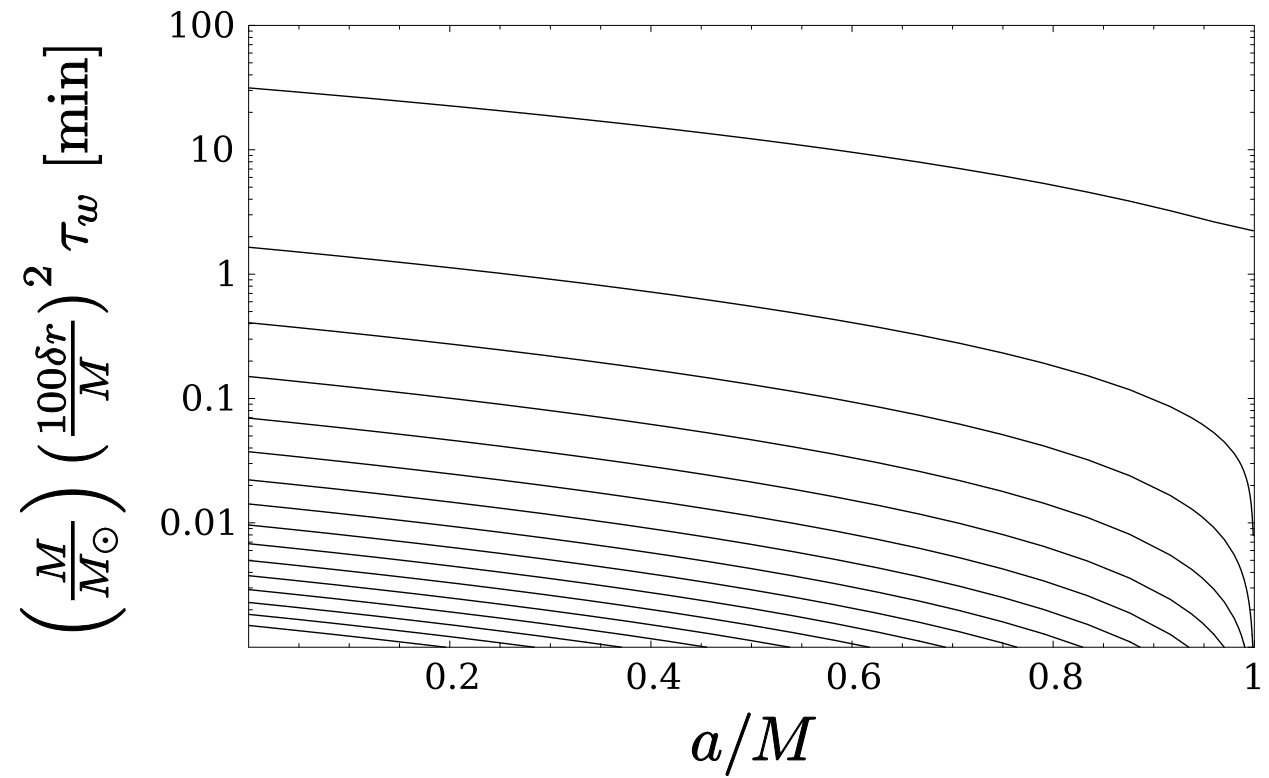}
\end{center}
\caption{The winding up timescale for spiral density waves (the
uppermost curve corresponds to $m=0$), giving an order of magnitude
estimate of their lifetime in minutes.  The lifetimes are much
longer than they need to be to produce the observed widths of the
QPO spectral peaks.} \label{fig:tauw}
\end{figure}
\clearpage

For completeness we present in Figure \ref{fig:tauw} the lifetimes
of various modes to winding up.  The lifetimes must be at least 10
times the inverse frequency in order that mode decay not broaden the
peaks in the power spectrum to a lower $Q$ value than is observed.
However, this condition is very easily satisfied for modes
concentrated near the discrete radii $r_m$.  The broadening of the
timing peaks is probably not due to finite mode lifetime unless the
modes extend in radius a significant fraction of the distance to the
next discrete mode radius $r_m$.

In addition to the beating of a blob with a wave, one might consider
the nonlinear interaction between waves of different pattern speeds.
However, this leads nowhere because the pattern (or modulation)
speed is a function of radius, and spiral density waves of different
mode number have different frequencies only because they are
localized at different radii.

The application to Kerr has until now considered only planar
perturbations. Vertical perturbations can also be excited and are
long-lived when concentrated at the discrete radii $r_n$ such that
$\nu(r_n)=n$. For the Kerr metric there are only two solutions
outside the ISCO:

\begin{enumerate}
\item For $a/M>0.9524$ there exists a solution with $n=0$ at
$r_0\approx1.9M$.  The frequencies are $f_\theta=\Omega_m/2\pi=4.9$
kHz, with a winding-up timescale of about $\tau =2$ min for $\delta
r\sim 0.01M$.
\item For $a/M>0.9987$ there exists a second solution with $n=-1$ at
$r_{-1}\approx1.2M$. The frequencies are $f_\theta=2.7$ kHz,
$\Omega_m/2\pi=16.5$ kHz, with a winding-up timescale of about $\tau
=0.4$ min for $\delta r\sim0.01M$.
\end{enumerate}

Because these modes occur outside the ISCO only for $a/M$ close to
maximal, the mode radii and frequencies are nearly constant over the
small range of black hole spin where they exist.  Given the very
narrow range of spins for which these long-lived warps exist, and
given their very high frequencies, they are likely to be
observationally unimportant.

At larger radii and for smaller black hole spins, the lifetime of
vertical waves becomes relatively large even though $\mu$ and $\nu$
are not integers.  To order of magnitude the winding time multiplied
by mode frequency is $\sim\mu\omega_m\tau_w\sim(r/\delta r)$ and
similarly for the vertical modes.  Also, the horizontal and
modulation frequencies become comparable: for the same mode number,
$0.9\lesssim\omega_m/\Omega_m<1$ for radii corresponding to pattern
frequencies in the range $10^{-4}$ Hz to $10^3$ Hz (multiplied by
$M_\odot/M$).  This frequency range includes the low-frequency QPOs
seen in the soft power-law state of black hole binaries
\citep{Remillard2006}.  The QPO frequency varies with X-ray flux,
implying that the modulation does not arise at a fixed radius
(unlike the high-frequency QPOs).

One explanation for the low-fequency QPOs is Lense-Thirring
precession (Stella et al.\ 1999, Schnittman et al.\ 2006), with
modulation speed $\Omega_{LT}=\omega_\phi-\omega_\theta$.  At large
radii, $\nu(r)\to1$, so vertical density waves with $n=1$ are
long-lived and have pattern speed equal to $\Omega_{LT}$. However,
planar perturbations have $\mu(r)\to1$ and are comparably
long-lived.  Therefore, either planar (spiral waves) or vertical
waves (warps) are candidate explanations for low-frequency QPOs, and
they have practically the same frequencies and lifetimes. Instead of
having no density wave solution consistent with observations (the
situation for high-frequency QPOs), the low-frequency QPOs have a
plethora of kinematic solutions with little to distinguish them.

\section{Conclusions}

Starting with a simple kinematical consideration --- fluid
streamlines should not intersect --- we were able to classify all
long-lived linear perturbations in pressureless thin accretion disks
in generic potentials including the Kerr metric.  The result is a
set of patterns (planar spiral arms, vertical warps, and linear
combinations) whose mode functions are concentrated at discrete
radii satisfying a simple condition: the pattern frequency for
non--self-intersecting orbits must be approximately constant over a
range of radii, i.e.\ its derivative with respect to radius must
vanish. This condition automatically picks out discrete set of
frequencies for modes with a long lifetime against winding up.  The
lifetimes are $O(r/\delta r)^2$ oscillation periods where $\delta r$
is the radial extent of the mode.

Although our treatment neglected pressure, the results are in
excellent agreement with the diskoseismic $g$-modes investigated by
\cite{Perez1997}, when the sound speed is much less than the orbital
speed.  Our treatment, based on the epicyclic approximation, is much
simpler than the solution of the relativistic fluid equations.

One of our goals was to investigate the possibility of explaining
the $3:2$ ratio in high-frequency quasi-periodic oscillations
observed in accreting black hole binaries. Our model gives several
mode pairs with a frequency ratio close to 1.5 for a wide range of
black hole spin. This result is a direct consequence of the discrete
radii at which the spiral patterns exist.  However, we cannot
recover naturally the correct frequencies unless the black hole mass
estimates are seriously in error.  Planar spiral density waves have
frequencies too high, while the beat frequency between a blob and
these waves is too low, compared with the observed frequencies.
Moreover, the best matches follow for modes with $|m|=2$ and
$|m|=3$, with no explanation for the failure to observe the mode
with $|m|=1$.

Because our analysis classified all long-lived perturbations of
pressureless disks, the failure to reproduce the observed
high-frequency quasi-periodic oscillations implies that pressure or
other non-gravitational forces (e.g.\ magnetic stresses or radiation
pressure) must be important in producing the QPOs.

Based on the diskoseismic results, we would expect that small
non-gravitational forces do not significantly change the mode
frequencies.  Large non-gravitational forces would thicken the disk
and might destroy the coherence and longevity of modes, although
this remains to be seen. Even so, some mechanism is needed to pick
out discrete frequencies of the high-frequency QPOs.

These results make more attractive the possibility that nonlinear
coupling between modes, e.g.\ parametric resonance, is responsible
for selecting the frequency ratios (Abramowicz \& Klu{\'z}niak 2001;
Rebusco 2004).  One approach to investigating this process is to
carry the perturbation methods of the current paper to one higher
order, to include a coupling between radial and vertical
oscillations.

\acknowledgements

This work was supported by NASA grant NAG5-13306.

\end{document}